\documentclass[aps,prl,twocolumn,superscriptaddress]{revtex4}
\usepackage{graphicx}
\usepackage{dcolumn}
\usepackage{bm}
\usepackage{color}

\begin{document}
%%%%%%%%%%%%%%%%%%%%%%%%%%%%%%%%%%%%%%%%%%%%%%%%%%%%%%%%%%%%%%%%%%%%%

% QUANTUM MECHANICS

\newcommand{\ket}[1] {\mbox{$ \vert #1 \rangle $}}
\newcommand{\bra}[1] {\mbox{$ \langle #1 \vert $}}
\def\vac{\ket{0}} \def\vacin{\ket{0}_{in}} \def\vacout{\ket{0}_{out}}
\def\thermal{\ket{\beta}}
\def\bvac{\bra{0}}\def\bvacin{{}_{in}\bra{0}}\def\bvacout{{}_{out}\bra{0}}
\def\bthermal{\bra{\beta}}
\newcommand{\ave}[1] {\mbox{$ \langle #1 \rangle $}}
\newcommand{\avew}[1] {\mbox{$ \langle #1 \rangle $}_w}
\newcommand{\vacave}[1] {\mbox{$ \bvac #1 \vac $}}
\newcommand{\thermalave}[1] {\mbox{$ \bthermal #1 \thermal $}}
\newcommand{\scal}[2]{\mbox{$ \langle #1 \vert #2 \rangle $}}
\newcommand{\expect}[3] {\mbox{$ \bra{#1} #2 \ket{#3} $}}
\def\a{\hat{a}}\def\A{\hat{A}}
\def\b{\hat{b}}\def\B{\hat{B}}
\def\aa{\tilde{a}}\def\AA{\tilde{A}}
\def\bb{\tilde{b}}\def\BB{\tilde{B}}
\def\kplus{\ket{+}}\def\kmoins{\ket{-}}
\def\bplus{\bra{+}}\def\bmoins{\bra{-}}

% GREEK LETTERS AND DIFFERENTIATIONS

\def\p{\prime}
\def\t{\tau}
\def\om{\omega}\def\Om{\Omega}
\def\ga{\gamma}
\def\omp{\om^\p}\def\Omp{\Om^\p}
\def\la{\lambda}\def\lap{\lambda^\p}
\def\mup{\mu^\p}\def\lp{l^\p}
\def\kp{k^\p}\def\sig{\sigma}
\def\ka{\kappa}
\def\al{\alpha}\def\alb{\bar\alpha}
\def\bt{\beta}\def\btb{\bar\beta}
\def\e{\epsilon}
\def\psip{\stackrel{.}{\psi}}\def\fp{\stackrel{.}{f}}
\def\lab{{\bar \la}}
\def\ffi{\varphi}
\def\scry{{\cal J}}
\def\scryp{{\cal J}^+}
\def\scrym{{\cal J}^-}
\def\scrypL{{\cal J}^+_L}
\def\scrymR{{\cal J}^-_R}
\def\scrypR{{\cal J}^+_R}
\def\scrymL{{\cal J}^-_L}
\def\Pig{{\mathbf \Pi}}

\def\di{\partial}
\def\diU{\di_U}
\def\diUU{\di_{\bar U}}
\def\diV{\di_V}
\def\diVV{\di_{\bar V}}
\def\didiv{\raise 0.1mm \hbox{$\stackrel{\leftrightarrow}{\di_V}$}}
\newcommand{\didi}[1]{\raise 0.1mm \hbox{$\stackrel{\leftrightarrow}{\di_{#1}}$}}

\def\d{\mbox{d}}
\def\D{\mbox{D}}

% LATEX WRITING

\def\TUU{T_{UU}}
\def\TUUI{T_{UU}^I}\def\TUUII{T_{UU}^{II}}
\def\Tuu{T_{uu}}
\def\TVV{T_{VV}}
\def\TVVI{T_{VV}^I}\def\TVVII{T_{VV}^{II}}
\def\Tvv{T_{vv}}
\def\re{\mbox{Re}}
\def\im{\mbox{Im}}
\def\S{{\mathbf S}}
\def\T{{\mathbf T}}
\def\1{{\mathbf 1}}
\def\Ss{\mbox{$\hat S$}}
\def\f{{\tilde f}}
\def\ubl{\bar u_L}
\def\UU{{\bar U}}
\def\VV{{\bar V}}
\def\LL{{\cal L_{\rm int}}}
\def\disp{\displaystyle}
\def\bitem{\begin{itemize}}
\def\eitem{\end{itemize}}
\def\bes{\begin{description}}
\def\es{\end{description}}
\newcommand{\be} {\begin{equation}}
\newcommand{\ee} {\end{equation}}
\newcommand{\ba} {\begin{eqnarray}}
\newcommand{\ea} {\end{eqnarray}}
\newcommand{\bsub}{\begin{subeqnarray}}
\newcommand{\esub}{\end{subeqnarray}}
\newcommand{\bwt} {\begin{widetext}}
\newcommand{\ewt} {\end{widetext}}

\def\cf{{\it cf}~}
\def\ie{{\it i.e.}~}
\def\etc{{\it etc}}
\def\eg{{\it e.g.}~}
\def\apriori{{\it a priori}~}

\def\nn{\nonumber \\}
\newcommand{\reff}[1]{Eq.(\ref{#1})}

\newcommand{\encadre}[1]{\begin{tabular}{|c|}\hline{#1}\\ \hline  \end{tabular}}

% MATH
\def\sinc{\mbox{sinc}}
\def\sinhc{\mbox{sinhc}}
\def\arccosh{\mbox{arccosh}}
\def\arcsinh{\mbox{arcsinh}}
\def\arctanh{\mbox{arctanh}}
\def\l{\left}
\def\r{\right}
%\int_{-\infty}^{+\infty}\! \mbox{d} #1 \;
\def\Si{\mbox{Si}}
\def\half{{1 \over 2}}
\newcommand{\inv}[1]{\frac{1}{#1}}
\def\inte{\int_{-\infty}^{+\infty}}
\def\into{\int_{0}^{\infty}}
\newcommand\Ie[1]{\inte \! \mbox{d} #1 \;}
\newcommand\Io[1]{\into \! \mbox{d} #1 \;}
\newcommand\IeIe[2]{\int \!\!\! \inte \! \mbox{d} #1 \: \mbox{d} #2 \;}
\newcommand\IoIo[2]{\int \!\!\! \into \! \mbox{d} #1 \: \mbox{d} #2 \;}
\newcommand\Iomoins[1]{\int_{-\infty}^{0} \! \mbox{d} #1 \;}
\newcommand\Iinfsup[3]{\int_{#1}^{#2} \! \mbox{d} #3 \;}
\newcommand{\erf}{\mathop{\rm erf}\nolimits}

\newcommand\lrpartial[1]{\mathrel{\partial_{#1}\kern-1em\raise1.75ex\hbox{$\leftrightarrow$}}}

\newcommand{\doublesum}[4]{\sum_{\begin{array}{cc}\scriptstyle #2 \\ \scriptstyle #1
\end{array}}^{\begin{array}{cc}\scriptstyle #4 \\ \scriptstyle #3 \end{array}}}

%Compteur de sous-equations
\newcounter{subequation}[equation] \makeatletter
\expandafter\let\expandafter\reset@font\csname reset@font\endcsname
\newenvironment{subeqnarray}
  {\arraycolsep1pt
    \def\@eqnnum\stepcounter##1{\stepcounter{subequation}{\reset@font\rm
      (\theequation\alph{subequation})}}\eqnarray}%
  {\endeqnarray\stepcounter{equation}}
\makeatother

\newcommand{\exergue}[2]{\begin{flushright}{\it\scriptsize #1} {\bf\scriptsize #2}\end{flushright}}

%%%%%%%%%%%%%%%%%%%%%%%%%%%%%%%%%%%%%%%%%%%%%%%%%%%%%%%%%%%%%%%%%%%%%%%%%%%%%%%%%%%%%%%%%%
%%%%%%%%%%%%%%%%%%%%%%%%%%%%%%%%%%%%%%%%%%%%%%%%%%%%%%%%%%%%%%%%%%%%%%%%%%%%%%%%%%%%%%%%%%
%%%%%%%%%%%%%%%%%%%%%%%%%%%%%%%%%%%%%%%%%%%%%%%%%%%%%%%%%%%%%%%%%%%%%%%%%%%%%%%%%%%%%%%%%%
%%%%%%%%%%%%%%%%%%%%%%%%%%%%%%%%%%%%%%%%%%%%%%%%%%%%%%%%%%%%%%%%%%%%%%%%%%%%%%%%%%%%%%%%%%

\title{How hot are expanding universes ?}
\author{Nathaniel Obadia}
\affiliation{\'Ecole Normale Sup\'erieure de Lyon, CRAL, UMR 5574 CNRS, 46 all\'ee d'Italie, 69364 Lyon Cedex 07, France }
\begin{abstract}
%A way to address the conundrum of Quantum Gravity is to build up connections between
%quantum fields techniques and General Relativity issues on the basis of some sound physical concepts.
%For intance, in order to illustrate the potentially fundamental interplay
%between quantum field theory, curved space-times physics and thermodynamics,
%one usually looks for the thermal properties of moving quantum systems in the vacuum of some specific space-times.
%So far, besides Hawking's black hole expression, the only known temperatures are those obtained by
%Unruh - uniform linear acceleration in Minkowski space-time,
%Gibbons and Hawking - inertial motion in de Sitter space-time
%and Narnhofer, Peter and Thirring - uniform linear acceleration in de Sitter space-time.
%These three cases require constant (or null) accelerations $A$ and Hubble parameters $H$.
A way to address the conundrum of Quantum Gravity is to illustrate the potentially fundamental interplay
between quantum field theory, curved space-times physics and thermodynamics.
So far, when studying moving quantum systems in the vacuum,
the only known perfectly thermal temperatures are those obtained
for constant (or null) accelerations $A$ in constant (or null) Hubble parameters $H$ space-times.
In this Letter,
restricting ourselves to conformally coupled scalar fields,
we present the most comprehensive expression for the temperature
undergone by a moving observer in the vacuum, valid for any time-dependent linear accelerations and Hubble parameters:
$T=\sqrt{A^2 + H^2 + 2 \dot H\dot t}/{2\pi}$
where $\dot t=\d t/\d\t$ is the motion's Lorentz factor.
The inequivalence between a constant $T$ and actual thermality is explained.
As a byproduct, all the Friedman universes for which observers at rest
feel the vacuum as a thermal bath are listed.
\end{abstract}
\maketitle
{\it Introduction}--
There are many ways to try to address the yet unknown issue of Quantum Gravity.
One of them is to focus on physical situations that naturally involve quantum physics
in curved space-time, like black holes (BH) or inflationary cosmology.
The quantum treatment of these issues' outcome provides strong similarities
with the basic laws of thermodynamics, mainly by assigning them a fixed temperature
and, if any, a given entropy.
This Letter attempts to go one step further in the study of thermodynamics applied to
quantum mechanics in curved space-times.

Following Bekenstein's work on BH entropy\cite{Bekenstein},
Hawking was the first one to assign an intrinsic temperature to a curved space-time issue
by showing that BH quantum evaporation in the Minkowski vacuum is thermal\cite{Hawking} with $T_{H}=\kappa/2\pi$,
$\kappa=1/4M$ is the BH's {\it constant} surface gravity ($G=k_B=\hbar=1$ henceforth).
Understanding the analogy between BH and cosmological event horizons,
Gibbons and Hawking\cite{GibbonsHawking} proposed that a similar temperature should be associated to de Sitter space-time too,
$T_{GH}=H_0/2\pi$, where $H_0$ is the {\it constant} Hubble parameter.
Transposing these intricate concepts to the safer ground of Minkowski space-time by invoking the equivalence principle,
Unruh\cite{Unruh} showed that a similar temperature is attributed to uniformly accelerated world-lines in Minkowski space-time,
$T_U=A_0/2\pi$, where $A_0$ is the world-line {\it constant} acceleration.
Relating these two issues, Narnhofer {\it et al.}\cite{Narnhofer} demonstrated that such world-lines also exhibit a temperature
in de Sitter space-time, $T_{N}=\sqrt{A_0^2+H_0^2}/2\pi$, followed by Deser and Levin\cite{DeserLevin} who found the corresponding expression
in the anti-de Sitter case.
Several generalizations of these results have been achieved in the literature:
While thermality in de Sitter space-time has been studied in \cite{Prokopec}
for massive and minimally coupled quanta in higher dimensions,
Jacobson\cite{Jacobson} replaced in the broader language of bifurcate Killing horizons Narnhofer {\it et al.}'s formula,
and the latter has been generalized in \cite{Buchholz} by introducing the concept of local temperatures.\\
\indent The aim of this Letter is to broaden the mapping
between thermodynamics and quantum field theory in curved space-times by
proposing a general formula for temperatures
of {\it time-dependent} $A[\t]$ accelerated linear world-lines
in {\it time-dependent} $H(t)$ Friedman-Lema\^{\i}tre-Robertson-Walker space-times
for conformally coupled scalar fields.
The method applied throughout this paper is the following:
we focus on the two-point Wightman function $W$ and identify thermality to when $W$ takes a special form.
From that, we infer a general formula for the temperature in the general case.
We then explore known and new situations where thermality is obtained.
\\
\indent{\it Friedman-Lema\^{\i}tre-Robertson-Walker space-times}--
Consider a FLRW universe with a flat metric
\ba\label{FLRW}
\d s^2=\d t^2-a^2(t)\: [\d x^2+\d y^2+\d z^2] \ .
\ea
The conformal time is defined by $\d \eta = \d t/a(t)$
such that, with the conformal coordinates $(\eta,\vec{x})$, the metric reads
$\tilde g_{\mu\nu}=a^2(t(\eta)) \: \mbox{diag}(1,-1,-1,-1)$.
%Along this paper $'=\di_t$.
Defining the Hubble parameter as $H=\di_ta/a$ one gets
\ba\label{eta}
\eta(t)=\Iinfsup{t_1}{t}{l} a(l)^{-1} + \eta_1 \; , \; a(t) = a_1 e^{\Iinfsup{t_1}{t}{l} H(l)}
\ea
where the subscript ${}_1$ denotes some reference time.
The only non-vanishing affine connections are
%$\Gamma^0_{i i}=H(t) e^{2\Iinfsup{}{t}{t'}H(t')}$ and $\Gamma^i_{i 0}=\Gamma^i_{0 i}=H(t)$.
$\Gamma^0_{i i}\! =\! H(t) a(t)^2$ and $\Gamma^i_{i 0}\! =\! \Gamma^i_{0 i}\! =\! H(t)$.
The Ricci tensor is diagonal
$R_{tt}\! =\! 3H^2+3\di_tH,R_{xx}\! =\! R_{yy}\! =\! R_{zz}\! =\! -(3H^2+\di_tH)a(t)^2$ and
the curvature scalar is $R=6(2H^2+\di_tH)$.
In the sequel we shall use the property that
no time-dependent $H$ FLRW space-time can be obtained by a coordinate transformation
from Minkowski ($H\!=\!0$) or de Sitter ($H\!=\!H_0$)\footnote{This comes from the fact that maximally symmetric spaces
in $3+1$ dimensions exhibit\cite{MisnerThorneWheeler}
$R_{\mu\nu}=(R/4)g_{\mu\nu}$,
which is only achieved by Minkowski and de Sitter among FLRW flat space-times.}.\\
\indent{\it Wightman functions}--
The Lagrangian for a conformally coupled massless scalar field in a curved
space-time background is given by
\ba
\sqrt{-g}{\cal L} = -\sqrt{-g}\left( \half g^{\mu\nu}(\di_\mu\Phi)(\di_\nu\Phi) -\inv{12}R\Phi^2 \right)\ .
\ea
where $g=\mbox{det}g_{\mu\nu}$.
When using the rescaled field $\varphi=a \Phi$ one gets the d'Alembert equation
$\left(\di_\eta^2 - \Delta\right)\varphi=0$
which enables to obtain the second quantized {\it physical} field
\ba\label{Phi}
\Phi(t,\vec{x})&=& \inv{a(t)}\:\Iinfsup{}{}{\vec{k}}%\inte \! \mbox{d} \vec{k} %\frac{\mbox{d} \vec{k}}{\sqrt{2{(2\pi)}^3k}} \;
\left(
b_{\vec{k}} \: \phi_{\vec{k}} %\frac{e^{-i (k \eta(t) - \vec{k}\vec{x})}}{\sqrt{2{(2\pi)}^3k}} \right.\nn
%&& \ + \
%\left.
+ b_{\vec{k}}^\dagger \: \phi_{\vec{k}}^* %\frac{e^{+i(k \eta(t) - \vec{k}\vec{x})}}{\sqrt{2{(2\pi)}^3k}}
\right) \\
\phi_{\vec{k}} &=& %\frac{e^{-i (k \eta(t) - \vec{k}\vec{x})}}{\sqrt{2{(2\pi)}^3k}}
{e^{-i (k \eta(t) - \vec{k}\vec{x})}}/{\sqrt{2{(2\pi)}^3k}}
\ea
where $k = |\vec{k}|$; $b_{\vec{k}}^\dagger$ and $b_{\vec{k}}$ are the creation and annihilation operators
obeying the usual commutation rules
$\left[ b_{\vec{k}},b_{\vec{k'}}^\dagger\right] = \delta(\vec{k}-\vec{k'})$.
A key-tool to study the physics of the physical field $\Phi$
is the $2$-point Wightman function
\ba \label{W1}
W(t,\vec{x};t',\vec{x'})&=&\vacave{\Phi(t,\vec{x})\Phi(t',\vec{x'})} \\
&=& \Iinfsup{}{}{\vec{k}} \: \frac{e^{-i[k(\eta(t)-\eta(t'))-\vec{k}.(\vec{x}-\vec{x'})]}}{16 \pi^3 k a(t)a(t')} \label{W2} \ .
\ea
This expression diverges as such in the ultra-violet and necessitates
regularization\footnote{More than its regularizability, the mere existence of the formula (\ref{W2}) comes from
the joined conditions of a) using a FLRW geometry in which constant time slices are conformally flat and
b) restricting ourselves to conformally coupled fields.}.
Along this article the field will be compelled to lie along given time-like world-lines noted $x^\mu[\t]$,
where $\t$ is the proper-time along the trajectory.
When labelling both events $(t,\vec{x};t',\vec{x'})$ by their proper-times $\t$ and $\t'$,
a convenient procedure to regularize \reff{W2} consists in shifting the labels in the complex plane according to \cite{OMi}
$\t\to \t-i\xi,\t'\to \t'+i\xi$;
$\xi>0$ being a small parameter sent to zero at the end of the calculation.
%If the Wightman function has to be integrated over in order to compute physical observables, t
The presence and the sign of $i\xi$ are mandatory to obtain the following regular results:
\ba
W[\t,\t']&=&-\Big[4\pi^2
\big\{ (\eta(t[\t-i\xi])-\eta(t[\t'+i\xi]))^2\\
&& \!\!\!\!\!\!\!\!\!\!- (\vec{x}[\t-i\xi]-\vec{x}[\t'+i\xi])^2  \big\} \: a(t[\t]) \: a(t[\t'])\Big]^{-1} \ .\nonumber
\ea
When restricting ourselves to linear trajectories $x^\mu[\t]=(t[\t],0,0,z[\t])$, the Wightman function reads
\ba
W[\t,\t']&=&\frac{-1}{4\pi^2} \left[
\left( \Iinfsup{t[\t'+i\xi]}{t[\t-i\xi]}{u} P[u;\t,\t'] \right)^2 \right. \label{Wgeneral} \\
&-& \left.
\left( \Iinfsup{t[\t'+i\xi]}{t[\t-i\xi]}{u} \beta(u) P[u;\t,\t'] \right)^2 \right]^{-1}  \nn
&=&\frac{-1}{4\pi^2} \left[
\left( \Iinfsup{-(\delta-i\xi)}{\delta-i\xi}{\la} \dot t[\bar\t+\la] Q[\la;\bar\t,\delta] \right)^2 \right. \label{Wgeneraltau} \\
&-&\left.
\left( \Iinfsup{-(\delta-i\xi)}{\delta-i\xi}{\la} \sqrt{\dot t[\bar\t+\la]^2-1}\;
Q[\la;\bar\t,\delta] \right)^2  \right]^{-1} \nn
P[u;\t,\t']&=& e^{ -\half \Iinfsup{t[\t]}{u}{l}H(l) -\half \Iinfsup{t[\t']}{u}{l}H(l)} \nn
Q[\la;\bar\t,\delta]&=& e^{ -\half \Iinfsup{\delta}{\la}{\sigma}\dot t[\bar\t+\sigma]H(t[\bar\t+\sigma]) -\half \Iinfsup{-\delta}{\la}{\sigma}\dot t[\bar\t+\sigma]H(t[\bar\t+\sigma])}
\nonumber
\ea
where $\dot{} = \di_\t$, $\delta=(\t-\t')/2$, $\bar \t=(\t+\t')/2$
and
$0 \leq \beta(t[\t]) = \frac{\sqrt{\dot t^2-1}}{\dot t} < 1 \ .$
The Wightman function only depends on the values of $H$ and of the trajectory
between $\t'$ and $\t$: this is the signature of causality.
%Consequently all the constants of integration appearing with the subscript ${}_1$ in \reff{eta}
%have disappeared from (\ref{Wgeneral}) and (\ref{Wgeneraltau}).
Note also that the integrals do not vanish when the two events coincide, thanks to the
regulator $i\xi$:
regulating the ultra-violet divergence for the modes in \reff{W2}
is equivalent to avoiding the small lapse singularity in
Eqs.(\ref{Wgeneral}) and (\ref{Wgeneraltau}).
Moreover, although the definition (\ref{W1}) trivially ensures $W[\t,\t']=W[\t',\t]^*$ with or without regulator,
the latter causes the inequality $W[\t,\t']\neq W[\t',\t]$ which encodes the fact the vacuum contains only positive
frequency modes.
Finally, \reff{Wgeneraltau}
%stresses that
%the effect of the regulator reduces merely to a shift in the lower half of the complex plane $\delta\to\delta-i\xi$.
shows that the regulation reduces to the mere shift $\delta\to\delta-i\xi$.
Keeping this in mind, we omit $i\xi$ in the sequel for the sake of clarity.\\
\indent From Eqs.(\ref{Wgeneral}) and (\ref{Wgeneraltau}), the Wightman function is written as a functional of $H(t)$
and $\dot t[\t]$, since $\dot z = e^{-\Iinfsup{}{t}{l}H(l)}\sqrt{\dot t^2-1}$.
More useful is to consider the acceleration along the world-line which square is
\ba\label{defacc}
A^2[\t]&=&-g_{\mu\nu}\frac{\D \dot x^\mu}{\D\t}\frac{\D \dot x^\nu}{\D\t}
\ea
where $\D$ is the covariant derivative for any $4-$vector $B^\mu$
%\ba\label{defcovariant}
$\frac{\D B^\mu}{\D\t}\equiv\frac{\d B^\mu}{\d\t} + \Gamma^\mu_{\al\beta} B^\al \dot x^\beta$
%\ea
and $g_{\mu\nu}\dot x^\mu \dot x^\nu=1$.
For linear trajectories one obtains
\ba\label{acc2}
A[\t] = {\ddot t}/{\sqrt{\dot t^2 - 1}} + H(t) \sqrt{\dot t^2 - 1}
\ea
when $\dot t \neq 1$; and $A=0$ when $\dot t=1$.
The acceleration in FLRW space-times comes from two sources:
the first term in \reff{acc2} is the extrinsic part inherent to the sole trajectory's acceleration
whereas the second one is the intrinsic part due to the universe's expansion.
The special case $\dot t=\gamma=cst>1$,
that we call ``gliding'' trajectories because their acceleration comes only from the universe's one,
possess the very simple $A[\t]=H(\gamma \t)\sqrt{\gamma^2 - 1}$.\\
\indent{\it The Minkowski case/Thermality}--
When studying linear trajectories in Minkowski space-time,
the Wightman function's Taylor development around $\delta=0$
involves the acceleration $A_M[\t]=\ddot t/\sqrt{\dot t^2 - 1}$ according to\cite{OMi}
\ba
W[\bar \t+\delta,\bar \t-\delta]&=&-\left(4\pi^2d^2[\bar \t+\delta,\bar \t-\delta]\right)^{-1} \label{Wd2}\\
d^2[\bar \t+\delta,\bar \t-\delta]&=&4\delta^2+\frac{4}{3}\delta^4 \ A_M^2[\bar\t]+ \mathcal{O}(\delta^6) \label{d2}
\ea
where $d$ is the geodesic distance between both events
and where the last term in \reff{d2} is {\it a priori} also a function of $\bar \t$.
Although it is not trivially deduced from the previous equations,
the condition $A_M=cst$ is {\it equivalent} to $W[\delta]$ only.
In other words, inertial ($\dot t=\gamma\geq 1, A_M=0$) and uniformly accelerated ($\dot t=\cosh[A_0\t], A_M=A_0$) linear trajectories are the only
stationary linear world-lines in Minkowski space\cite{Letaw}:
\ba
W_M^{in}[\bar \t+\delta,\bar \t-\delta]&=&- \left( 4 \pi \delta \right)^{-2} \label{WMinkIn} \\
W_M^{u.acc}[\bar \t+\delta,\bar \t-\delta;A_0]&=&- \left( 4 \pi \sinh\left[A_0\delta \right]/A_0  \right)^{-2} \label{WMinkUA} \ .
\ea
\reff{WMinkUA} happens to be also the Wightman function along an inertial world-line in a thermal bath at temperature
\ba
T_U={A_{0}}/{2\pi} \label{TMinkUA} \  .
\ea
This result, the Unruh effect\cite{Unruh},
induces that uniformly accelerated observers in the Minkowski vacuum feel as embedded in a thermal bath.
It is crucial to notice that the thermality and the value of the temperature itself
are univocally defined by the functional structure of the Wightman function (\ref{WMinkUA}),
\ie its stationary character and its hyperbolic sine functional dependence in $\delta$.
Though useful tools, two-level detectors\cite{OMi}, mirrors\cite{OPa2,OPa3} or other quantum devices\cite{Crispino}
are not mandatory to reveal thermality: {\it Thermality is equivalent to a Wightman function that can be written
as \reff{WMinkUA}}.\\
\indent{\it The general case}--
Using \reff{Wgeneraltau}, one finds that, for any linear trajectory in any FLRW universe,
\ba
&&W[\bar \t+\delta,\bar \t-\delta]=-\left(4\pi^2D[\bar \t+\delta,\bar \t-\delta]\right)^{-1}\label{Wmostgeneral}\\
&&D[\bar \t+\delta,\bar \t-\delta]=4\delta^2+\frac{4}{3}\delta^4 \ B[\bar\t]+ \mathcal{O}(\delta^6) \label{d2v3}\\
&&B[\t] = A^2 + H^2 + 2 \di_tH\dot t^2 = A^2 + H^2 + 2 \dot H\dot t  \label{Aeff}\ .
\ea
{\it A priori} $B$ is neither constant nor even positive.
%Therefore a) a uniform $B$ is a {\it necessary} condition for stationary solutions,
%and b) if one looks for thermality, \ie for the hyperbolic sine dependence,
%$B$ should be positive.
%In this case, its square root defines a temperature as in Eqs.(\ref{d2}) and (\ref{TMinkUA}).
If one requires a {\it stationary} Wightman function, a uniform $B$ is a {\it necessary} condition.
Moreover, if one requires a {\it thermal} Wightman function, \ie the hyperbolic sine dependence (\ref{WMinkUA}),
a positive $B$ is another {\it necessary} condition.
Therefore, thermality implies that the temperature should take the value $\sqrt{B}/2\pi$ as $A_M/2\pi=A_0/2\pi$
in Eqs.(\ref{d2}) and (\ref{TMinkUA}).
Hence the following theorem:\\
\indent {\it In a flat FLRW space-time given by its Hubble parameter $H(t)$,
if a linearly moving observer with acceleration $A[\t]$ feels the vacuum
as a thermal bath, the Wightman function along the world-line is equal to\footnote{One could ask why thermality cannot be achieved
by adding in the r.h.s of \reff{theoremW} a suitable ``reminder'' (that would not alter the thermal response
function of a detector, for instance). By subtracting \reff{theoremW} to \reff{Wmostgeneral}, one finds that
this reminder should be at least $\mathcal{O}(\delta^2)$. However, if one wishes to recover a Planck ratio for 
the excited and ground levels of a detector moving along the trajectory, 
the reminder should be at least $\mathcal{O}((\delta-i\e)^{-4})$\cite{OMi}. 
Therefore, no reminder can be added and thermality has to be defined by the strict equality \reff{theoremW}.}
\ba\label{theoremW}
W[\bar \t+\delta,\bar \t-\delta]&=&- \left( 4 \pi \sinh\left[2\pi T_{eff}\delta \right]/2\pi T_{eff}  \right)^{-2}
\ea
and the corresponding (constant) temperature is\footnote{As it should be,
\reff{theorem} possesses a covariant rewriting:
$H$ and $\di_tH$ are functions of $R$ and $R_{\mu\nu}R^{\mu\nu}$, and $\dot t=\dot H/\di_tH$.}}
%by the constant and well-defined
\ba\label{theorem}
T_{eff}= {\sqrt{A^2+H^2+2\di_tH \dot t^2}}/{2\pi} \ .
\ea
%Next I fathom this theorem in the well-known de Sitter case as well as for other space-times
%and then I show that the reciprocal implication ``$T_{eff}=cst\!\Rightarrow\!$ thermality'' is not fulfilled,
%contrary to the Minkowski case.\\
Let us fathom this result in the well-known de Sitter case and then extend its applicability to other space-times. \\
\indent {\it The de Sitter case}--
When considering de Sitter space-time $(H(t)=H_0)$,
$T_{eff}=cst$ implies either inertial or uniformly accelerated linear trajectories.
Inertial trajectories are either
the static solution $\dot t = 1$
or the extrinsically counter-accelerating one $\dot t = 1/\tanh[H_0(\t-\t_0)]$,
defined for $\t>\t_0$.
Considering these two types of inertial world-lines, one gets
\ba\label{WdSIn}
W_{dS}^{in}[\t,\t';H_0]=- \left( 4 \pi \sinh\left[H_0\delta\right]/H_0  \right)^{-2} \ .
\ea
The latter expression is exactly the Wightman function
obtained for uniformly accelerating trajectories in Minkowski space-time with $A_M=H_0$ (\ref{WMinkUA}).
The equivalence between both situations finds its roots in the fact that $3+1$ de Sitter
can be seen as a time-like hyperboloid, parametrized by its radius $1/H_0$, embedded in $4+1$ Minkowski space-time.
The geodesics in the first one are homomorphic to uniformly accelerated world-lines in the second one with $A_M=H_0$.
From this analogy, one recovers the Gibbons-Hawking's temperature\cite{GibbonsHawking}
\ba
T_{dS}^{in} = T_{eff} = %\frac{H_0}{2\pi}\label{TdSIn}\ .
{H_0}/{2\pi}\label{TdSIn}\ .
\ea
Uniformly accelerated world-lines in de Sitter are obtained, for example,
with the gliding trajectories $(A_G=H_0\sqrt{\gamma^2-1})$.
The Wightman function is also thermal
\ba\label{WdSUA}
W_{dS}^{u.acc}[\t,\t';H_0;\gamma]&=&-
\left( 4 \pi \sinh\left[H_0\gamma\delta\right]/(H_0\gamma)  \right)^{-2}
\ea
with the Narnhofer {\it et al.} temperature\cite{Narnhofer}
\ba
T_{dS}^{u.acc} = T_{eff} = %\frac{H_{0}\gamma }{2\pi} = \frac{\sqrt{A_G^2+H_0^2}}{2\pi}  \label{TdSUA} \ .
{H_{0}\gamma }/{2\pi} = {\sqrt{A_G^2+H_0^2}}/{2\pi}  \label{TdSUA} \ .
\ea
\indent {\it Other thermal cases}--
Our method to look for other solutions is to impose the necessary condition $B[\t]=cst\geq 0$
and to fix one degree of freedom (either the cosmology $H(t)$ or the trajectory $\dot t[\t]$).
Once a possible solution is found, if the Wightman obeys \reff{theoremW},
one deduces thermality with the temperature (\ref{theorem}).\\
\indent First let us look for cases for which the vacuum appears as an effective one, \ie $T_{eff}=0$.
Apart from the trivial case (\ref{WMinkIn}),
\reff{Aeff} requires the Hubble parameter to be a strictly decreasing function of time
and to satisfy
\ba
B[\t] &\equiv& (H^2 + 2 \di_tH)\dot t^2 +2H \ddot t + \frac{\ddot t^2}{\dot t^2-1} = 0\ .
\ea
Therefore all the gliding trajectories $t=\gamma\t$, including the inertial ones ($\gamma=1$),
have $A[\t]=2\sqrt{\gamma^2-1}/\t$, and
possess a vanishing effective temperature {\it iff} $H^2 + 2 \di_tH=0 \Leftrightarrow H(t)=H_W(t)=2/(t-t_0)$.
This Hubble parameter is a specific example of power-law inflation\cite{PLI} with the
equation of state $\om=-2/3$ and emerges as a solution for a wall-dominated phase\cite{Seckel}.
Moreover, if one restricts oneself to $t>t_1\gtrsim t_0$, $H_W$ is a viable model
of non-slow-rolling transition from a large Hubble factor $H_W(t_1)$ universe to an asymptotically vanishing curvature one.
One can easily check with \reff{Wgeneraltau} that $t=\gamma\t$ gives back
the same Wightman function as in the Minkowski/inertial case (\ref{WMinkIn}):
when $H(t)=H_W(t)$,
inertial trajectories ($A=0$), as well as gliding ones ($A\propto 1/\t$), feel the vacuum as an effective one
as they do in Minkowski space
\ba\label{H2sur3}
H_W(t)= 2/(t-t_0) \Rightarrow T_W^{in} = T_W^{G} = T_{eff} = 0 \ .
\ea
Note however that gliding trajectories are accelerated in this universe contrary to Minkowski space.
Remark also that, though Wightman functions {\it for these trajectories} are equivalent in both space-times,
Minkowski and $H_W$-FLRW universes differ since no coordinate transformation can bring one to the other.
In the sequel we give an example that explicitly splits the degeneracy between both space-times.\\
\indent Contrary to the previous example, requiring non-zero temperatures
allows also non-decreasing Hubble parameters solutions.
Focusing on inertial and gliding trajectories,  $\dot B=2\gamma^3(H \di_tH+\di_t^2H)=0$
has for only non-trivial solutions the family of functions
\ba\label{HZ}
H_P(t)=H_0 \tanh(H_0(t-t_0)/2)
\ea
where $t_0$ and $H_0$ are free parameters.
Since $H_P$ is always increasing it is a typical example of phantom cosmology\cite{Caldwell}
for which the equation of state is time-dependent $\om = -1 -\sinh^2[H_0(t-t_0)/2]/2 < -1$.
It describes a contracting and bouncing back space-time that possesses de Sitter as an attractor
and can be obtained with a conformally coupled massive inflaton with a $\lambda \varphi^4$ potential\cite{HZ}.
This model could be used, for instance, to describe the onset of near-de Sitter inflation
by some other mechanism.
Plugging this solution into \reff{Wgeneral} one obtains a similar result as previously
for inertial and gliding trajectories though this time involving an equivalence with de Sitter
\ba\label{WHZin}
W_{H_P}^{in}[\t,\t';H_0]&=&W_{dS}^{in}[\t ,\t';H_0]  \\
W_{H_P}^{G}[\t,\t';H_0,t_0;\gamma]&=&W_{dS}^{u.acc}[\t,\t';H_0;\gamma] \label{WHZacc}
\ea
which means that, although the Hubble parameter depends on time,
inertial and gliding observers in a $H_P(t;H_0,t_0)$ cosmology feel the vacuum as
if they were imbedded in a $H_0$ de Sitter space:
\ba
T_{H_P}^{in} = T_{eff} =%\frac{H_0}{2\pi} \; , \;
{H_0}/{2\pi} \; , \;
T_{H_P}^G = T_{eff} =%\frac{H_0\gamma}{2\pi} \ .
{H_0\gamma}/{2\pi} \ .\label{THp}
\ea
Since in $H_P$ space gliding trajectories do not possess a uniform acceleration as they do in de Sitter,
the similarity between both space-times is not perfect, once again.\\
\indent As a byproduct of the last paragraphs,
one learns that, using the subgroup of $\dot t=1$ inertial world-lines,
two other FLRW universes possess an intrinsic temperature
(\ie related to inertial trajectories) besides Minkowski and de Sitter:
$H_W(t)$ with $T^{in}_{eff}=0$,
and $H_P(t;H_0)$ with $T^{in}_{eff}=H_0/2\pi$.
Considering $\dot t\neq 1$ inertial world-lines,
can other space-times be intrinsically thermal ?
The reparametrization $\dot t[\t]=1/\tanh(\alpha[\t])>1$ in \reff{acc2},
provides all the other solutions
\ba
H[\t]=\dot\alpha \ , \ T_{eff}^{in}=\sqrt{\dot\alpha^2+2\ddot\alpha/\tanh\alpha}/2\pi \ . \label{intrinsic}
\ea
Injecting the latter quantities in \reff{Wgeneraltau} and imposing $B=4\pi^2 T_{eff}^2=cst$,
the $\delta^6$ term in \reff{d2v3} is $(8/45)(B^2-(B-\dot\alpha^2)^2/4\cosh^2\alpha)$
instead of the $8B^2/45$ an hyperbolic sine provides:
Minkowski $(H=\dot\alpha=\sqrt{B}=0)$ and
de Sitter $(H=\dot\alpha=\sqrt{B}\neq 0)$ with the trajectory quoted above \reff{WdSIn}
are the only solutions.
{\it Hence the only intrinsically thermal FLRW space-times are Minkowski, $H_W$, de Sitter and $H_P$.}\\
\indent{\it Necessary but not sufficient}--
The trajectory $\dot t[\t]=\cosh[A_0\t]$, which provides thermality in Minkowski space-time,
possesses a uniform acceleration $A=3A_0$ and a constant $T_{eff}=\sqrt{5}A_0/2\pi$ when $H=H_W$.
However, the corresponding Wightman function is not stationary and not thermal.
The latter result shows that, though necessary, $T_{eff}=cst$ is not sufficient to claim for thermality;
this should not be a surprise.
In \cite{OMi}, it has been shown that only in the particular case of {\it linear} trajectories in {\it Minkowski} space-time,
a constant acceleration is equivalent to a thermal Wightman function.
As soon as $2D$-exploring trajectories are concerned this equivalence is broken though Eqs(\ref{Wd2}) and (\ref{d2}) are still valid.
a) For instance see the cusp motion which exact expression generates the ``circular Unruh effect''\cite{Circular1,Circular2}:
there, a constant acceleration $A_0$ provides a stationary but {\it non-thermal} Wightman function.
When using two-level detectors, two quasi-temperatures are found according to the probed part of the spectrum,
but neither one nor the other equals $A_0/2\pi$.
b) More drastic than the former example, any uniform acceleration but time-dependent torsion planar world-line is non-stationary\cite{OMi}.
Therefore, adding one degree of freedom breaks the implication $T_{eff}=cst \Rightarrow$ {\it thermality},
whether $T_{eff}$ is a perfect or even a quasi-temperature.
With this respect, the expansion caused by the Hubble factor acts as an additional degree of freedom
and has the same side-effect.\\
\indent {\it Summary}--
In this Letter the concept of ``temperature of a moving observer in a curved space-time vacuum'' is extended
to all time-dependent Hubble parameters and linear accelerations for conformally coupled fields:
If the vacuum is felt as a thermal bath along a world-line in a given FLRW space-time, its temperature is $T_{eff}$
given in \reff{theorem}.
As a general law for such systems, {\it thermality} ${\mathbf T}$ always implies {\it stationarity} ${\mathbf S}$.
Here we found that, if ${\mathbf T}$ naturally requires $T_{eff}$ to be constant,
the inverse implication is not true.
This is rooted in the fact that, as was found for (hyper-)torsion for some non-linear(non-planar) world-lines\cite{Letaw,OMi},
there is more than one degree of freedom in the problem:
in our case, due to the non-linear differential character of the Wightman function that entangles $A,H,\etc$,
the curvature of space-time prevents ${\mathbf S} \Rightarrow {\mathbf T}$ to be always true.
Moreover, neither ${\mathbf S}$ nor ${\mathbf T}$  are equivalent to $\dot H=0$,
as is epitomized in the second result of this letter:
two $H(t)$ space-times, Eqs.(\ref{H2sur3}) and (\ref{HZ}), are
intrinsically thermal, \ie for observers at rest.
Though significatively time-dependent\footnote{Their first slow-roll parameters are
$\e_W=1/2$ and the unbounded $\e_P=-(\cosh(H_0(t-t_0))-1)^{-1}$.},
these space-times share their thermal properties with their respective attractors,
Minkowski and de Sitter,
thereby allowing an isothermal transition to them. %:
In order to interpret where isothermality is rooted,
one has to find the physical cause of similar Green functions
either by referring directly to the space-times Killing properties\cite{Jacobson}
or by invoking common causal horizons characteristics\cite{JacobsonParentani};
this is the subject of a work in progress\cite{Oinprogress}.
Another question this Letter raises is what happens for "real" universes,
for which the Hubble parameter is not exactly constant or equal to Eqs.(\ref{H2sur3}) and (\ref{HZ});
that is whether or not quasi-thermality is achievable for slowly varying $T_{eff}$
(as was done for near-uniform acceleration world-lines in Minkowski space\cite{OMi})\cite{Oinprogress}.
A better understanding of these issues will help to unveil the foundations of
Quantum Gravity by exploring the intricacy of its thermodynamics.\\
${\bf{Acknowledgements}}$
I thank M.Milgrom, B.D.Bruner and an anonymous referee for very helpful remarks as well
as the CIMeC, Trento U. for their hospitality.

\end{document}